\documentstyle[12pt]{article}
\def\abstract#1{\vskip 7mm 
	\begin{center}{\large Abstract}\par \bigskip
		\begin{minipage}[c]{12cm}
			\small #1
		\end{minipage}
	\end{center}
}
\def\title#1{\begin{center}{\Large\bf #1}\end{center}}
\def\author#1{\vskip 5mm \begin{center}{#1}\end{center}}
\def\address#1{\begin{center}{\it #1}\end{center}}

\newcommand{\bfr}{\begin{flushright}}
\newcommand{\efr}{\end{flushright}}

\begin{document}

\vspace*{-2cm}
\bfr{}\efr\vspace{-9mm}
\bfr{}\efr
\vspace{1cm}

\title{Role of the imaginary part in the Moyal quantization}
\author{Takao KOIKAWA\footnote{E-mail: koikawa@otsuma.ac.jp}}
\vspace{1cm}
\address{
  School of Social Information Studies,
         Otsuma Women's University,\\
         Tama 206-0035,Japan\\
}
\vspace{2.5cm}
\abstract{ 
We show that the imaginary part of the $\star$-genvalue equation in the Moyal quantization reveals the symmetries of the Hamiltonian by which we obtain the conserved quantities. Applying to the Toda lattice equation, we derive conserved quantities which are used as the independent variables of Wigner function.}

\newpage
\setcounter{page}{2}
Wigner functions have been receiving attention in various fields of physics. They were invented by Wigner and Sziland\cite{WigSzi}. Wigner functions underlie Moyal's formulation of quantum mechanics\cite{Moy}, and obey the $\star$-genvalue equation\cite{Curt,Fair} which consists of an imaginary part and a real part. 
In this letter, we study the role of the imaginary part of the $\star$-genvalue equation and show that it serves to extract the symmetries of the Hamiltonian in the phase-space and so that of Wigner function. We study this in a simple example of the harmonic oscillator first, and then the Toda lattice equation which is known as one of the soliton equations. 
The classical soliton equations are known to possess an infinite number of conserved quantities which are the generators of the transformations which leave the Hamiltonian invariant. We study the role of the conserved quantities in quantum mechanics of Moyal's formulation. 
We showed that the soliton equations can be formulated by the zero-curvature equation using the Moyal star product with a parameter $\kappa$, but in the present letter, the parameter is a different one $\hbar$ and so the discussion here is not directly relevant to the zero-curvature formulation of solitons\cite{Koi}.
 
In the phase-space, the Hamiltonian of the harmonic oscillator possesses the rotational symmetry. In Moyal's formulation we can find the same generator of the symmetry in the imaginary part. We can reduce the number of independent variables by using this symmetry, which helps to solve the real part of the $\star$-genvalue equation. 
From this lesson we learn that the imaginary part of $\star$-genvalue equation is to be used to find out the symmetries of the Hamiltonian. 
It is well known that the Toda lattice equation of $N$ particles with the periodic boundary condition has $N$ conserved quantities which were found by H$\grave {\rm e}$non\cite{Henon}.
It is interesting to apply the Moyal's formulation of quantum mechanics to the soliton equations which are known to have many, often an infinite number of symmetries. We apply the $\star$-genvalue equation to the Hamiltonian of the Toda lattice equation, find the symmetries of the Hamiltonian, and use the conserved quatities as variables of Wigner function. 

We first consider the Hamiltonian of the harmonic oscillator given by
\begin{equation}
H(q,p)=\frac{p^2}{2m}+k\frac{q^2}{2},
\end{equation}
where we assume $m=k=1$. By introducing a vector $v=(q,p)^t$, this is rewritten as
\begin{equation}
H(q,p)=\frac{1}{2}v^t v,
\end{equation}
where $v^t$ represents the transposed $v$. This is invariant under the infinitesimal transformation
\begin{equation}
v \to v'=(1+\theta {\hat R})v, 
\end{equation}
where $\theta$ is an infinitesimal parameter 
and the generator of the rotation $\hat R$ is given by
\begin{equation}
\hat R=q\frac{\partial}{\partial p}-p \frac{\partial}{\partial p}.
\label{eq:rotation}
\end{equation}
The invariance of the Hamiltonian shows the rotational symmetry in the classical phase-space.

The star product of functions $f=f(q,p)$ and $g=g(q,p)$ is defined by
\begin{equation}
f\star g = \exp \Bigg[ i\frac{\hbar}{2}
\Bigg(\frac{\partial~}{\partial q}\frac{\partial~}{\partial\tilde p}-\frac{
\partial~}{\partial p}\frac{\partial~}{\partial\tilde q}\Bigg)
\Bigg] f({\bf x}) g({\bf \tilde x}) \vert_{{\bf x} = {\bf\tilde x}},
\label{eq:star1}
\end{equation}
where ${{\bf x}=(q,p)}$ and ${{\bf \tilde x}=(\tilde q,\tilde p)}$  and they are set equal after the derivatives are taken. By using the star product, the $\star$-genvalue equation for the Hamiltonian $H(q,p)$ of the harmonic oscillator reads
\begin{equation}
H(q,p) \star f(q,p)=Ef(q,p),
\end{equation}
where f(q,p) is the static Wigner function. The real and imaginary part are respectively given by
\begin{eqnarray}
\frac{1}{2}\{ q^2+p^2-\frac{1}{4}(\frac{\partial^2}{\partial q^2}+\frac{\partial^2}{\partial p^2})\}f(q,p)&=&Ef(q,p),
\\
(q \frac{\partial}{\partial p}-p\frac{\partial}{\partial q})f(q,p)&=&0.
\end{eqnarray}
As is seen from the comparison of the imaginary part of the $\star$-genvalue equation and the generator $\hat R$ in Eq.(\ref{eq:rotation}), the rotational symmetry of the Hamiltonian and so Wigner function in the phase-space can be found from the imaginary part of the $\star$-genvalue equation. This symmetry enables one to solve the imaginary part equation to give:
\begin{equation}
f=f(z),
\end{equation}
where $z=2(q^2+p^2)$. The variable is proportional to the conserved quantity or the Hamiltonian itself. By using the variable $z$, the real part equation is rewritten and then we are left with solving the ordinary differential equation to obtain the $\star$-gen Wigner function. 
This shows that the conserved quantities are used as the variables of Wigner function in quantum mechanics.

Encouraged by this observation, we next consider one of the most famous soliton equations, or the Toda lattice equation. In the classical soliton equations, it is well known that there are conserved quantities more than 2. We are interested in how those symmetries are reflected in the quantum mechanics a la Moyal. We show that the symmetries of Wigner function appear in the Moyal's quantum formulation and they are identical to the classical ones. The Hamiltonian is given by
\begin{equation}
H=\frac{1}{2}\sum_n p_n^2+\sum_n\{e^{-(q_{n+1}-q_n)}-1\},
\end{equation}
where the summation is taken over the indices of $N$ particles. We also impose the periodic boundary condition. The $\star$-genvalue equation yields both the real and imaginary part as
\begin{eqnarray}
&&\frac{1}{2}\sum_n \{ p_n^2-\frac{\hbar^2}{4} \frac{\partial^2}{\partial q_n^2}+2e^{-(q_{n+1}-q_n)}\rm{cos} \frac{\hbar}{2}(\frac{\partial}{\partial p_{n+1}}-\frac{\partial}{\partial p_n})-2\}f(q,p)\nonumber\\ 
&=&Ef(q,p), 
\label{eq:TodaR}\\
&&\sum_n\{p_n\frac{\partial}{\partial q_n}+\frac{2}{\hbar}e^{-(q_{n+1}-q_n)}\rm{sin}\frac{\hbar}{2}(\frac{\partial}{\partial p_{n+1}}-\frac{\partial}{\partial p_n})\}f(q,p)=0.
\label{eq:TodaI}
\end{eqnarray}
We focus on the 2nd equation to derive the conserved quantities. Expanding the sin function we obtain
\begin{equation}
\sum_n\{p_n\frac{\partial}{\partial q_n}+{\rm e}^{-(q_{n+1}-q_n)}(\frac{\partial}{\partial p_{n+1}}-\frac{\partial}{\partial p_n})+O(\hbar^2) \}f(q,p)=0.
\end{equation}
Note that the rank of the partial derivatives with respect to each variable $p_i$ in $O(\hbar^2)$ is more than one. This means that, as far as we restrict $f(q,p)$ to be linear in each $p_i$ or $f(q,p) \propto \prod_i p_i$, this part is negligible. Actually we find such type of solutions. 
Then the remaining part is exactly the operator found in search of the conserved quantities of the classical Toda lattice equation by Sawada and Kotera\cite{SawaKote}. We can follow their discussion. The H$\grave {\rm e}$non-type solutions to (\ref{eq:TodaI}) are given by 
\begin{equation}
I_{N-n}=\big(\sum_j \frac{\partial}{\partial p_j}\big)^n \rm{e}^{-\rho} \prod_{i=1}^N p_i,(n=0,1,\cdots,N-1)
\end{equation}
where $\rho$ is given by
\begin{equation}
\rho=\sum {\rm e}^{-(q_{n+1}-q_n)}\frac{\partial ^2}{\partial p_n \partial p_{n+1}}.
\end{equation}
Thus we find that Wigner function is expressed as
\begin{equation}
f(q,p)=f(I_1,I_2,\cdots,I_N).
\end{equation}
We can set $N \to \infty$ or consider the Toda lattice equation with both ends at $\pm \infty$. In such cases, Wigner function is the function of an infinite number of variables. This shows that the quantum Toda lattice equation with these boundary conditions are described by Wigner function of an infinite number of variables. It is intriguing to note that the description of the soliton equations in quantum mechanics needs an infinite number of variables just as in the classical solutions to the soliton equations.

In summary we show that the imaginary part of the $\star$-genvalue equation reveals the symmetries of the Hamiltonian. Applying to the Toda lattice equation, we derive conserved quantities by which Wigner function is expressed. The remaining work is to rewrite Eq.(\ref{eq:TodaR}) in terms of the variables $I_{N-n}(n=0,1,\cdots,N-1)$ and solve it, which needs more investigation.

\end{document}